\documentclass[conference]{IEEEtran}
\IEEEoverridecommandlockouts

\usepackage[cmex10]{amsmath}
\usepackage{cite}
\usepackage{color}

\usepackage{amssymb}
\usepackage{path}
\usepackage{verbatim}
\usepackage{algorithm}
\usepackage{framed}
\usepackage{algpseudocode}
\usepackage{footnote}
\usepackage{color}
\usepackage{amsmath}
\usepackage{cases}
\usepackage{subfigure}
\usepackage{theorem}
\usepackage{subfloat}

\usepackage{tabularx}
\usepackage{booktabs}
\usepackage[colorlinks=true, allcolors=black]{hyperref}

\usepackage{graphicx}

\usepackage{pgfplots}
\usepackage{pgfplotstable}
\pgfplotsset{compat=newest}

{ \theorembodyfont{\normalfont} 

}

\newcounter{enumctr}

\DeclareFontFamily{U}{mathx}{\hyphenchar\font45}
\DeclareFontShape{U}{mathx}{m}{n}{<-> mathx10}{}
\DeclareSymbolFont{mathx}{U}{mathx}{m}{n}
\DeclareMathAccent{\widebar}{0}{mathx}{"73}

\newcolumntype{C}{>{\centering\arraybackslash}X} 
\begin{document}

\title{\LARGE \bf
	Parking Behaviour Analysis of Shared E-Bike Users Based on a Real-World Dataset - A Case Study in Dublin, Ireland}

\author{Sen Yan, Mingming Liu$^{*}$ and Noel E. O’Connor
	\thanks{S. Yan, M. Liu and N. E. O'Connor are with the SFI Insight Centre for Data Analytics and the School of Electronic Engineering, Dublin City University, Dublin, Ireland. This work is supported by Science Foundation Ireland SFI Grant SFI/12/RC/2289\_P2.}	
	\thanks{$^{\star}$Corresponding author. Email: {\tt mingming.liu@dcu.ie}}
}

\maketitle
\thispagestyle{empty}
\pagestyle{empty}

\begin{abstract}\label{sec:intro}
In recent years, an increasing number of shared E-bikes have been rolling out rapidly in our cities. It therefore becomes important to understand new behaviour patterns of the
cyclists in using these E-bikes as a foundation for the novel design of shared micromobility services as part of the realisation for next generation intelligent transportation systems. In this paper, we deeply investigate the users' behaviour of shared E-bikes in a case study by using the real-world dataset collected from the shared E-bike company, MOBY, which currently operates in Dublin, Ireland. More specifically, we look into the parking behaviours of users as we know that inappropriate parking of these bikes can not only increase the management costs of the company but also result in other users' inconveniences, especially in situations of battery shortage, which inevitably reduces the overall operational efficacy of these shared E-bikes. Our work has conducted analysis at both bike station and individual level in a fully anonymous and GDPR-Compliant manner, and our results have shown that up to 12.9\% of shared E-bike users did not park their bikes properly at the designated stands. Different visualisation tools have been applied to better illustrate our obtained results.
\end{abstract}

\begin{IEEEkeywords}
Shared E-Bikes, Parking Behaviour Analysis, Micromobility
\end{IEEEkeywords}

\IEEEpeerreviewmaketitle

\section{Introduction}

In recent years, electric bicycles (E-bikes) have been widely recognised as an effective and economical alternative to driving especially for people commuting short distances in our cities \cite{mcqueen2020bike, nocerino2016bikes}. By sharing the E-bikes among different users on demand,  i.e., shared E-bikes, the economical benefits of daily commuting for citizens can be improved, leading to a more flexible and efficient travel pattern, which forms an essential part of the Mobility-as-a-Service (MaaS) framework in the context of smart cities \cite{becker2020assessing, farahmand2021mobility}. To illustrate this point, the research reported in \cite{CAIRNS2017327} shows that the average distance of weekly commuting trips by E-bikes can range from 9.8km to 17km, and the trips using shared E-bikes can replace up to 76\% short trips that were made by private cars. Furthermore, in comparison with other transportation tools using fossil fuels, E-bikes are not only more environmentally friendly but also can give rise to health benefits to a wide range of road users, including both young and old people, as E-bikes require less physical energy input from riders thanks to the mechanism of electrical assistance \cite{gu2018design}. A recent study in \cite{KROESEN2017377} further indicates that E-bikes can generally motivate riders to cycle faster and over longer distances compared to traditional bikes, where the study points out that E-bike users can cycle 3.0km on average per day while the pedal bike riders can commute only around 2.6km on average in the same timeline. 

With the numerous benefits of E-bikes bringing to our lives, it is expected that an increasing number of shared E-bikes will be rolling out rapidly on roads in the near future. In fact, this already happens in Ireland as the Irish government has recently approved E-bike legislation to allow for these electrified bikes in road traffic in the same way as pedal cycles \footnote{https://www.gov.ie/en/press-release/12185-government-approves-next-steps-for-escooter-and-ebike-legislation/}. However, this new way of transportation will also create new challenges for traffic and operation management in our cities. More specifically, shared E-bikes, unlike pedal cycles, rely on available infrastructure to ensure their batteries are charged at the appropriate level in order to fulfil users' requests for their specific journeys. Therefore, it is essential that careful consideration must be given to these bikes to ensure that they can be parked, charged and managed properly in the desired manner, in a manner that all users can easily access to these bikes as they wish. 

To this end, a considerable research effort have been devoted in recent years globally to the analysis of parking behaviour of shared bikes/E-bikes users and the design of efficient parking management strategies for these tools \cite{su11195431, su11185003, 10.1007/978-3-030-14745-7_7, li_zhao_he_axhausen_2020, su11195439, 9564831}. For instance, in \cite{su11195431}, the authors pointed out some negative effects caused by disorderly parking in Beijing, China, and proposed some management suggestions by using a collaborative governance model to analyse parking behaviour of shared-bikes. In \cite{10.1007/978-3-030-14745-7_7}, a Mixed-Path Size logit model was devised to analyse the route choice decisions of E-bike users in the Netherlands. The work in \cite{su11185003} carried out analysis of both spatial and temporal characteristics of E-bike user' mobility in Tengzhou City, China. In \cite{li_zhao_he_axhausen_2020}, a trip purpose imputation framework was developed to assist in trip purpose prediction based on the trip records of different micro-mobility tools before and during the COVID-19 pandemic period in Zurich, Switzerland, and finally in \cite{su11195439}, a hierarchical clustering method was employed to address the issue of free-floating bike-sharing parking in Beijing, China. However, most of the works have been found either using mathematical models with certain assumptions on users' statistical characteristics of travelling or using realistic data but with a focus on the aggregated pattern. Few works have been found using realistic data but with a focus on the analysis for individual E-bike user, especially in relation to each user's parking behaviour, as reported in this paper. 

Our objective in this paper is to deeply investigate the parking behaviour of shared E-bike users at individual level by analysing the real-world dataset provided by MOBY Bikes \footnote{https://mobybikes.com/}, one of the main shared E-bikes providers based in Dublin, Ireland, as a case study. In particular, we wish to gain a deep understanding on how users parked their shared E-bikes under the current parking regulations of the company through comprehensive data analysis and visualisation, with an aim to reveal insight regarding how current parking strategies may be improved to encourage more users to better park their bikes for other users' convenience. 

\begin{figure}[ht]
	\vspace{-0.1in}
	\centering
	\subfigure[Bike with low battery]{
		\includegraphics[width=0.22\textwidth]{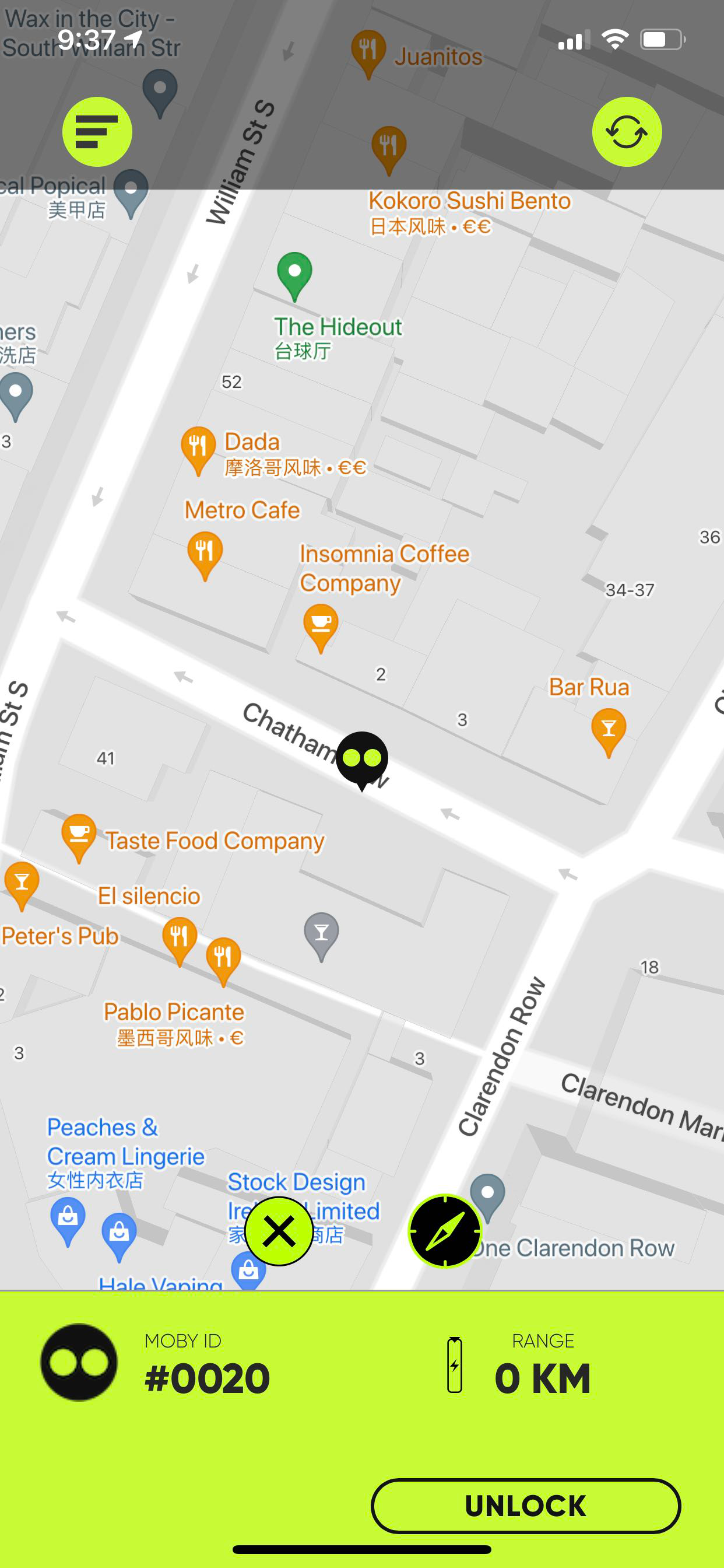}
		\label{subfig:fig1a}
	}
	\subfigure[Bike station with no bikes]{
		\includegraphics[width=0.22\textwidth]{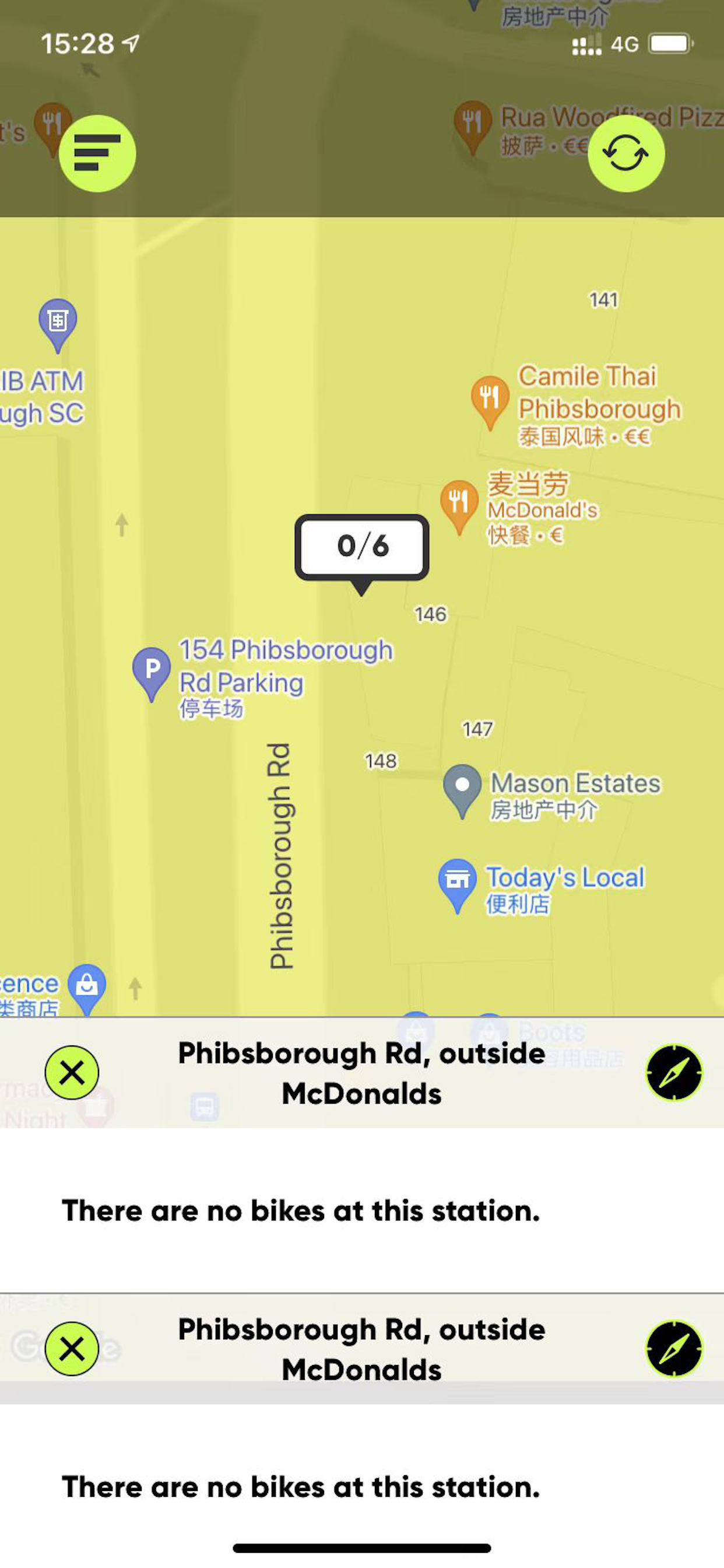}
		\label{subfig:fig1b}
	}
	\caption{Inconvenience when using the app.}
	\label{fig:inconvenience}
	\vspace{-0.1in}
\end{figure}

For a better context, according to the current regulation of MOBY, any disorderly parking of shared-bikes will incur a small monetary penalty and batteries in their shared E-bikes are required to be manually swapped by a team of operators when they are running low. Clearly, an inappropriately parked shared E-bike with a low battery pack can only be replaced depending on the availability of operators at the given time and space, this will inevitably increase the inconvenience for forthcoming users although a small penalty has to be paid by the previous rider. To further illustrate this point, two subplots from the MOBY App are shown in \autoref{fig:inconvenience} demonstrating the possibility of a potential forthcoming user with difficulties accessing an E-bike due to bikes with low battery for travel, shown in Figure \ref{subfig:fig1a}, and no bike currently available in the designated station as shown in Figure \ref{subfig:fig1b}. 

The rest of paper is organised as follows. In Section \ref{sec:data}, we describe the data obtained from MOBY, introduce the analysis of the dataset and demonstrate several preliminary analytical results based on this. In Section \ref{sec:user}, we look into user parking behaviours in detail. Finally, we conclude the paper in Section \ref{sec:conclusion} and outline some future research directions.

\section{Data Description and Analysis}\label{sec:data}

In this section, we describe, analyse and visualise the data collected by MOBY Bikes. It is important to highlight that the dataset received from the company consisting of both user rental reports and GPS reports is purely for academic research purpose as part of the Smart DCU \footnote{https://smartdublin.ie/smart-districts/smart-dcu/} programme and is provided in a fully anonymous and GDPR compliant manner.

\subsection{Dataset Description}
\begin{itemize}
    \item GPS Information: This part the dataset reflects the GPS coordinates for each bike at different moments with 5 features (i.e., \textit{Bike ID}, \textit{Latitude}, \textit{Longitude}, \textit{Valid} and \textit{Time}) for over 4 million records collected from January 2020 to September 2021. 
    
    \item Journey Records: This is the other part of dataset showing around 58,000 journey records with 37 features in total. Users are anonymous in the dataset in order to comply with GDPR. The feature set includes information of the user, E-bike (e.g., \textit{Bike ID}, \textit{Bike type}), consumption (e.g. \textit{Total price}, \textit{Commercial conditions}), time (e.g., \textit{Return time}, \textit{Rental duration}), position (e.g., \textit{Rental location}), and distance for each journey made by the user. Some briefly descriptive statistical information for total duration and distance of journey records are reported in \autoref{data_description}.
\end{itemize}

\begin{table*}
\caption{Summary of descriptive statistics for duration and distance for all E-bike journey records.}
\label{data_description}
\begin{tabularx}{\textwidth}{@{}l*{10}{C}c@{}}
\toprule
Features    & Minimum   & Maximum   & Median    & Mean  & Standard deviation\\ 
\midrule
Trip Distance (metres)  & 1.00  & 7914.00   & 371.00    & 586.66    & 653.90\\ 
Trip Duration (minutes) & 0.17  & 4587.18   &  21.62    &  56.53    & 131.93\\ 
\bottomrule
\end{tabularx}
\end{table*}

\subsection{Data Preprocessing}
\begin{itemize}
    \item GPS Information: When processing the GPS data, we found some duplicated records in the dataset. In particular, this relates to the feature \textit{Time}, where duplicated records were found for the same timestamp due to the received GPS signals from multiple satellites at the same time. To solve this, we replaced these coordinates to the mean value in the records for every given bike and timestamp. Some redundant features in the dataset were also discarded, one of which includes the feature \textit{Valid} which has the same value ``1'' in every record. Finally, after removing a small percentage of records with missing values, the cleaned dataset can be used to demonstrate the trajectory information of all bikes geographically.

    \item Journey Records: Some of the features in this dataset do not contain valuable information (i.e., \textit{Voucher code} and \textit{Status}) in relation to our analysis, so these columns were ignored in the data preprocessing step. A few records with value ``null'' in some features of interest, such as \textit{Distance} and \textit{Return time}, were removed for simplicity of our analysis. Finally, we also adopted the Z-score to determine and remove outliers from features with extraordinary high/low values in the dataset.
\end{itemize}

\subsection{Data Visualization}

\begin{figure}[ht]
    \vspace{-0.1in}
    \centering
    \includegraphics[width=0.5\textwidth]{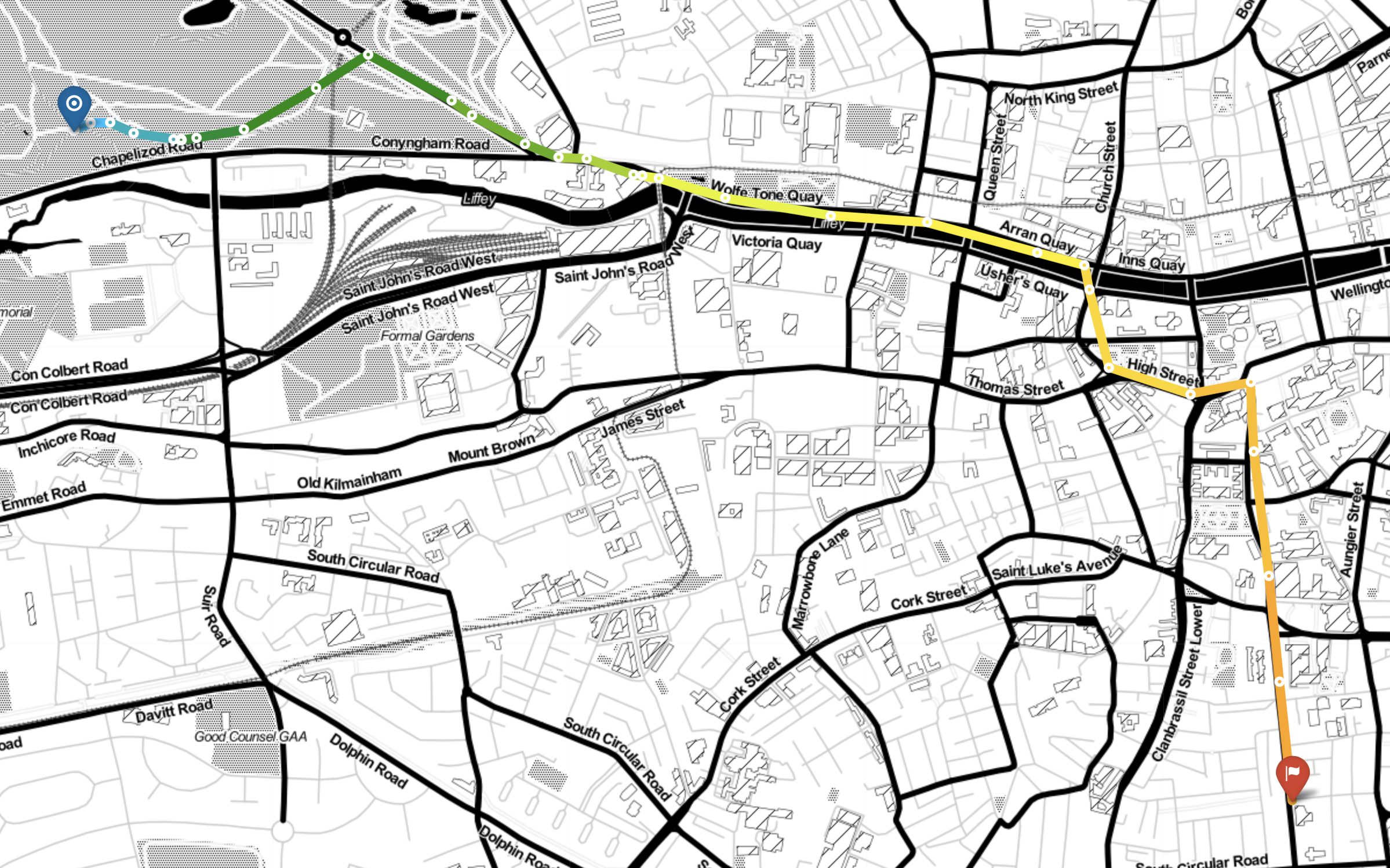}
    \caption{Trajectory of bike No.40 on a specific date.}
    \label{fig:fig2}
	\vspace{-0.1in}
\end{figure}

To illustrate the dataset, a trajectory of the Bike No.40 on a specific date is shown in \autoref{fig:fig2}, where the user started on the Chapelizod Road shown in blue, and terminated on the South Circular Road in Dublin 8 shown in red. 
    
\begin{figure}[ht]
    \vspace{-0.1in}
    \centering
    \includegraphics[width=0.5\textwidth]{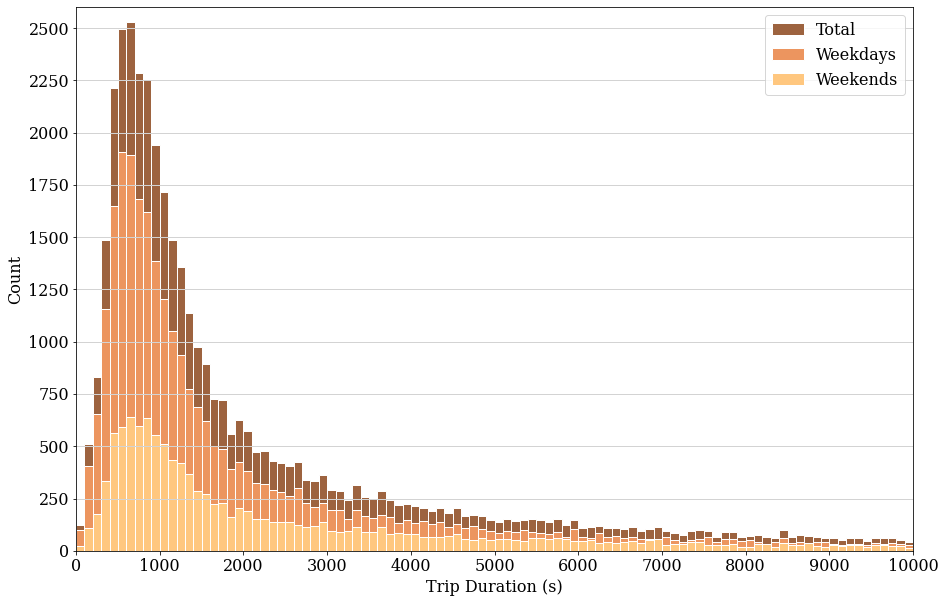}
    \caption{The histogram of trip duration using the MOBY dataset.}
    \label{fig:fig10}
	\vspace{-0.1in}
\end{figure}

In \autoref{fig:fig10}, the histogram of trip duration is demonstrated and compared between weekdays and weekends. It is shown that the trip duration for most users is between 600 seconds and 800 seconds, and in particular, the distribution of trip duration on weekdays is similar with that on weekends but with more number of trips being reported.

\begin{figure}[ht]
	\vspace{-0.1in}
	\centering
	\includegraphics[width=0.5\textwidth]{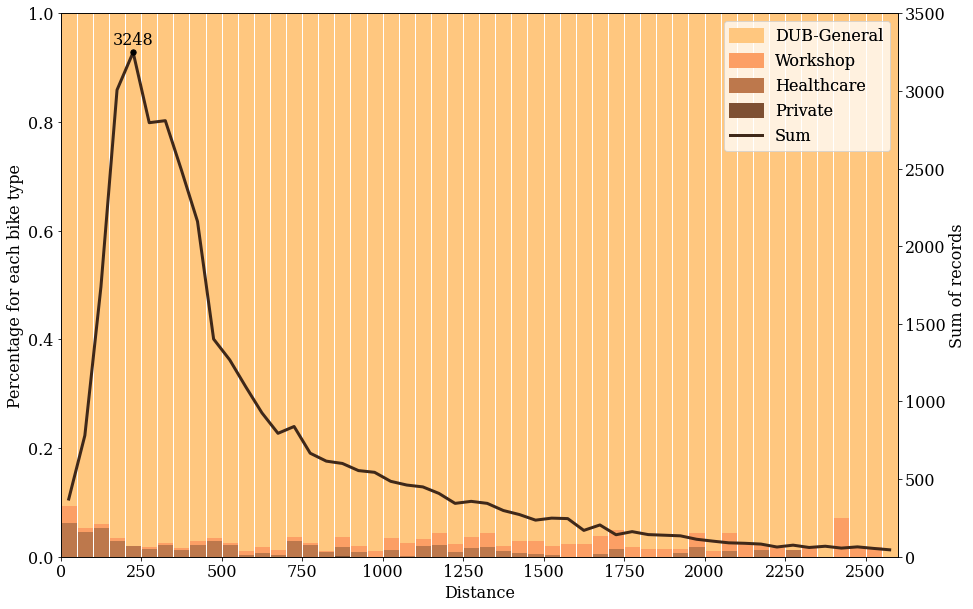}
	\caption{Percentage and sum of records for each E-bike type.}
	\label{fig:fig3}
	\vspace{-0.1in}
\end{figure}

Further, \autoref{fig:fig3} shows sum of records and the percentage values for each E-bike type with respect to different distance internals. It can be seen that the total number of travel records of all E-bikes, shown in the dark line in the figure, keeps increasing sharply till a point when it reaches to the peak level of 3,248 corresponding to the distance interval in the range of 200 to 250 metres. The number of records then falls back gradually to zero when distance interval is increased to a point just beyond 2500 metres. The \textit{DUB-General} type of E-bike users have clearly dominated the results with at least 90.59\% users in each distance internal especially in comparison to the \textit{Private} type of E-bike users which only appear once in the range of 850 to 900 metres.
    
\begin{figure}[ht]
	\vspace{-0.1in}
	\centering
	\includegraphics[width=0.5\textwidth, height=2.7in]{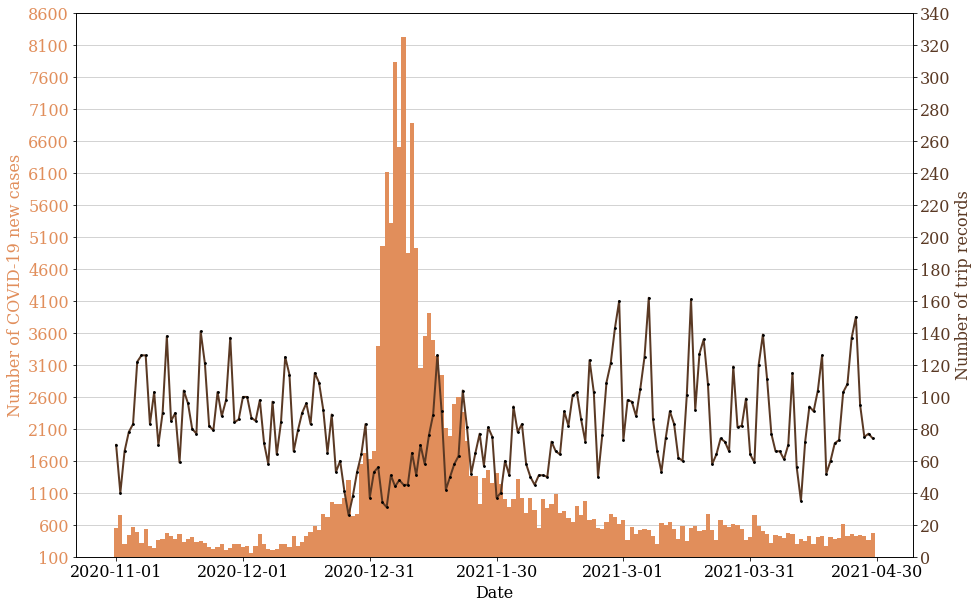}
	\caption{The number of trip records of MOBY bikes during the COVID-19.}
	\label{fig:fig4}
	\vspace{-0.1in}
\end{figure}

Since our dataset contains trip records during the COVID-19 pandemic periods, we also briefly analyse the impact of COVID-19 on the E-bike users' travelling behaviours. We do this by referring to the publicly available COVID-19 dataset for Dublin Ireland based on the COVID-19 Data Repository provided by the Center for Systems Science and Engineering (CSSE) at Johns Hopkins University \cite{dong_du_gardner_2020}. Our result is presented in \autoref{fig:fig4}, where we can see that the daily number of new cases in COVID-19 shown in yellow is negatively correlated with the number of trip records for MOBY bike users shown in black. However, we note that the rate of change for the number of records made by the shared E-bike users on average is significantly less in comparison to the rate of change in the COVID-related time-series data. To some extent, it demonstrates that sharing E-bike scheme is still attractive for people travelling in Ireland during COVID.
    
\section{Parking Behaviour Analysis} \label{sec:user}

In this section, we explore the shared E-bike users' parking behaviour in detail.

\subsection{Return Accuracy} \label{RA}

As aforementioned, inaccurate returns will significantly increase the difficulty of E-bike management and lead to inconvenience for other users. Thus, it is important to understand which users often return their bikes inconsiderably, and why this may happen. To do this, we need to understand the definition of an inappropriate return and then calculate the return accuracy for a given user. 

\begin{figure}[ht]
	\vspace{-0.1in}
	\centering
	\includegraphics[width=0.5\textwidth]{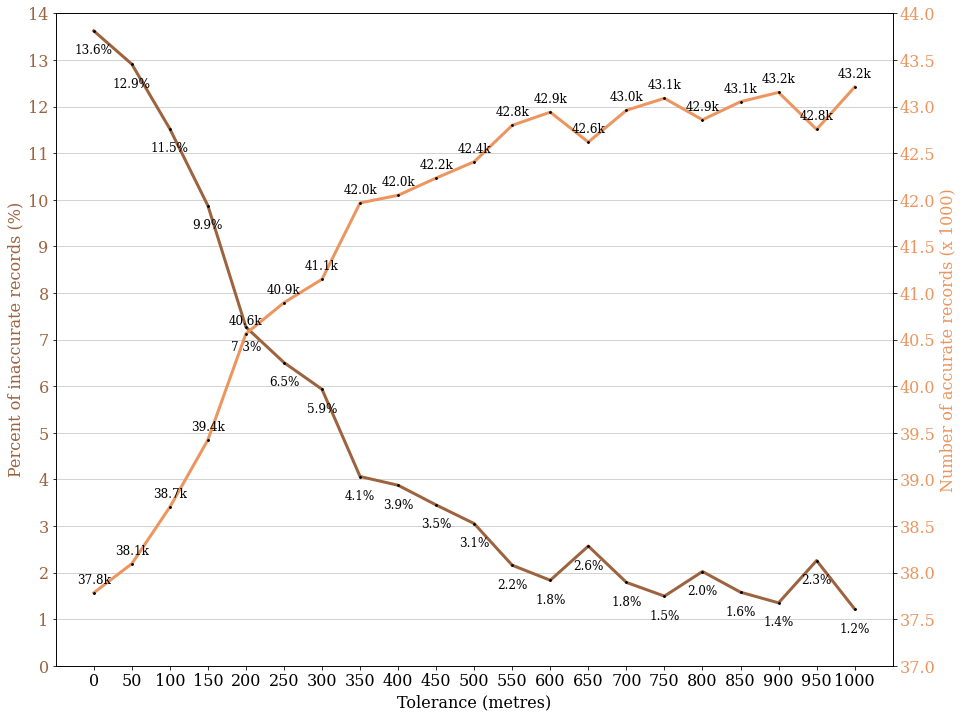}
	\caption{Sum of accurate and percentage of inaccurate returns.}
	\label{fig:fig5}
	\vspace{-0.1in}
\end{figure}

According to MOBY, the GPS signal from bike is accurate within a range of 10 to 15 metres, and it is possible for this range to be further increased to 50 metres in case there are buildings around. Thus, before calculating the return accuracy, we need to determine some high-use return positions with a clustering algorithm by setting a specific radius as tolerance to accommodate those potential mistakes made by GPS devices on the bikes. To do this, we applied the Canopy cluster algorithm, an unsupervised pre-clustering algorithm that has been widely used in data science \cite{Kumar2014CanopyCA}. More specifically, Canopy algorithm was applied to conduct a brief clustering for the return coordinates of bikes based on their GPS coordinates. The algorithm uses two thresholds, namely $T_{2}$ and $T_{1}$, to determine the minimum and maximum distance between two points to be clustered in the same group. If the exact distance between two points locates in the range of $(T_{2}, T_{1})$, then both points will be attached with the same cluster label, including both accurate stands predefined by MOBY and the inappropriate high-use positions we found using the algorithm. Thus, by setting $T_2 = T_1$ and varying the $T$ values, we can evaluate the return accuracy as a function of threshold, i.e., range tolerance. The results in \autoref{fig:fig5} illustrate that with an increasing level of range tolerance for GPS signals, the number of accurate returns is also increased gradually. According to this analysis, there were 12.9\% of MOBY users who did not park their shared-bikes properly with respect to the highest GPS tolerance level set by MOBY, i.e., 50 metres.

\begin{figure}[ht]
	\vspace{-0.1in}
	\centering
	\includegraphics[width=0.47\textwidth, height=2.2in]{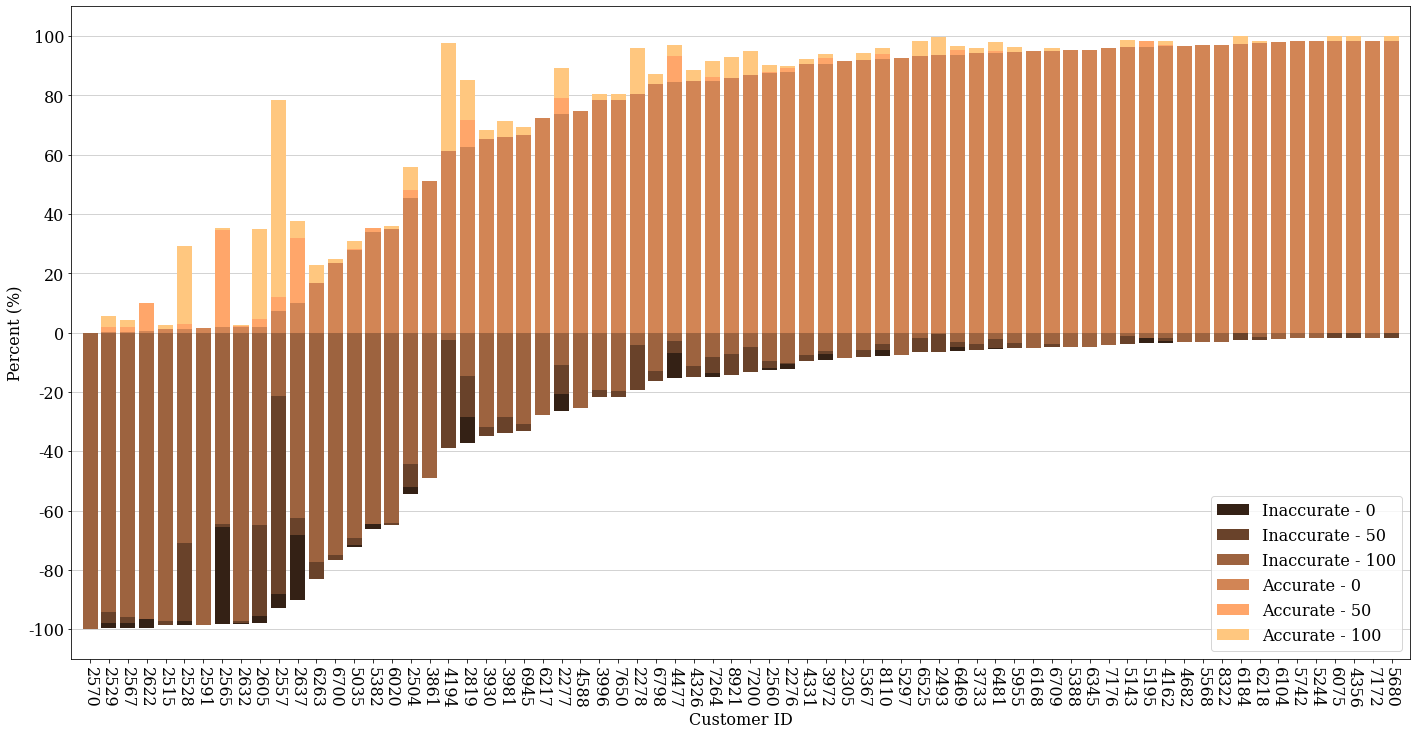}
	\caption{The impact of tolerance on user's return accuracy.}
	\label{fig:fig6}
	\vspace{-0.1in}
\end{figure}

After clustering, the return accuracy for each user can be easily calculated, which is shown in \autoref{fig:fig6}. In the figure, the bars in different colours represent the percentages of inaccurate and accurate returns with respect to three tolerance thresholds (i.e., 0, 50 and 100 metres). It could be seen that as the tolerance grows, the accuracy for some users (e.g. user 2565 and 2557) obviously improves. However, we also find that there are several customers who always returned their E-bike inaccurately even with tolerance of 50 metres (e.g., user 2570 and 2557).

\begin{figure}[ht]
	\vspace{-0.1in}
	\centering
	\includegraphics[width=0.48\textwidth, height=2.2in]{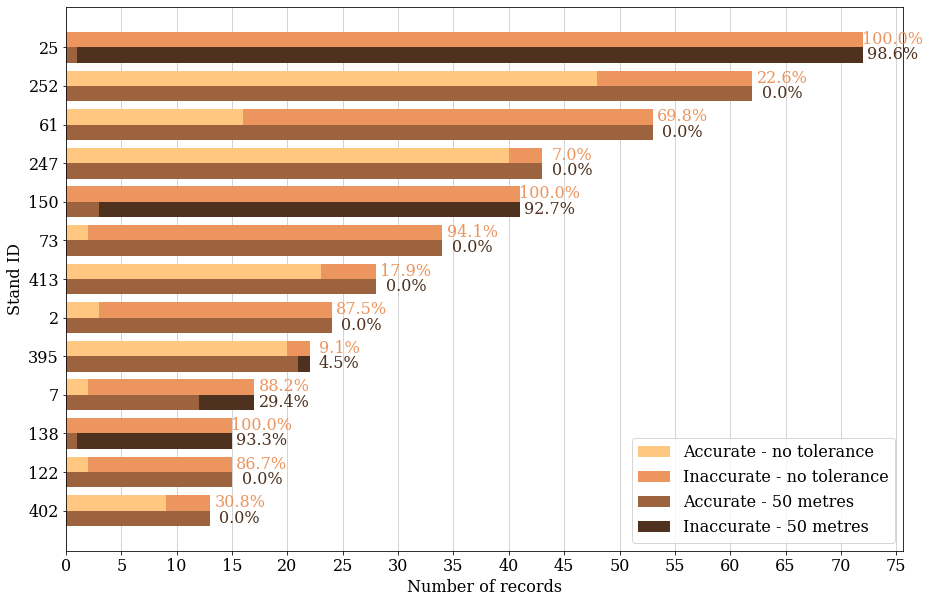}
	\caption{The impact of tolerance on return accuracy for stands.}
	\label{fig:fig7}
	\vspace{-0.1in}
\end{figure}

\begin{figure*}[ht]
	\vspace{-0.1in}
	\centering
	\subfigure[Stand 25]{
		\includegraphics[width=0.3\textwidth]{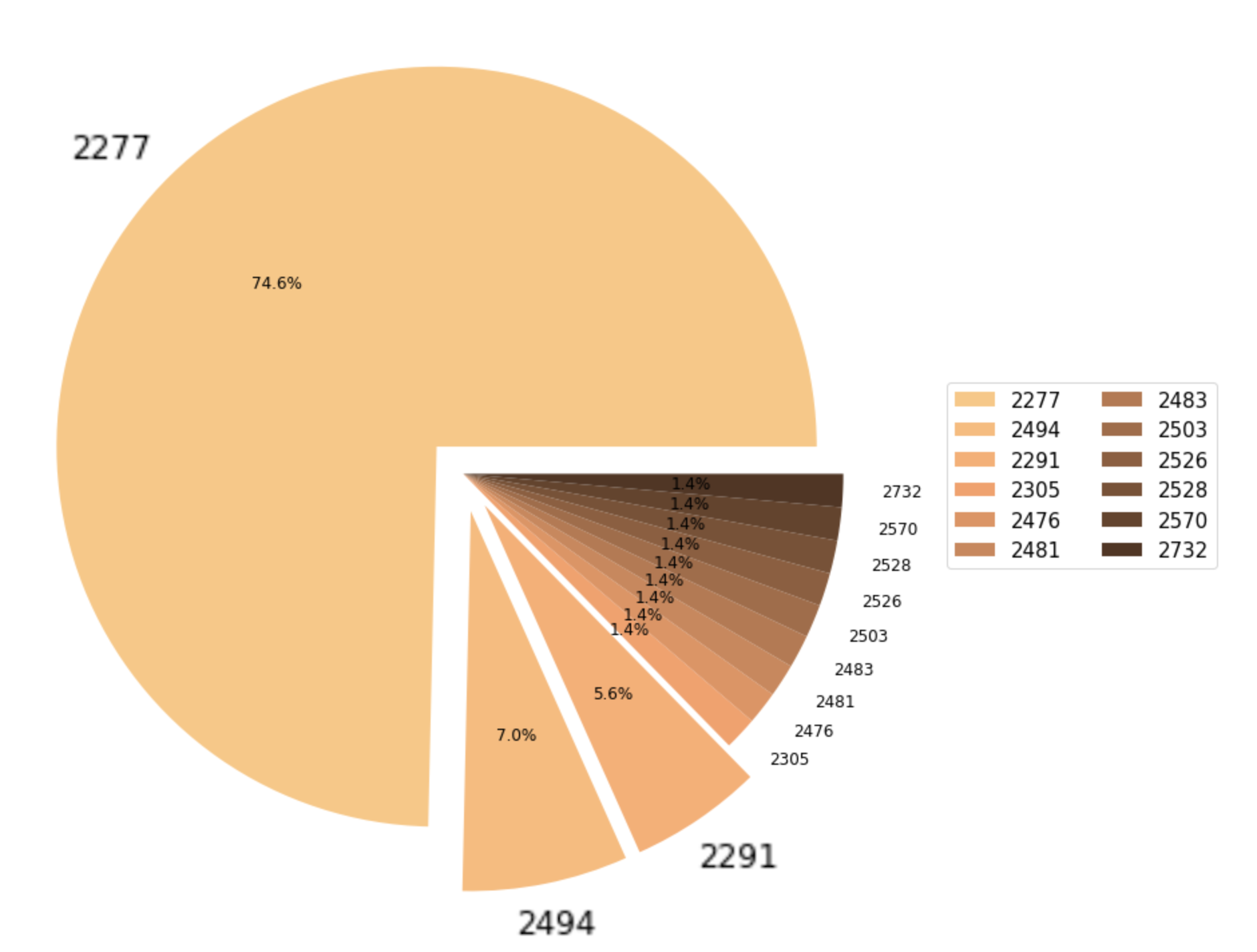}
		\label{subfig:s25}
	}
	\subfigure[Stand 150]{
		\includegraphics[width=0.3\textwidth]{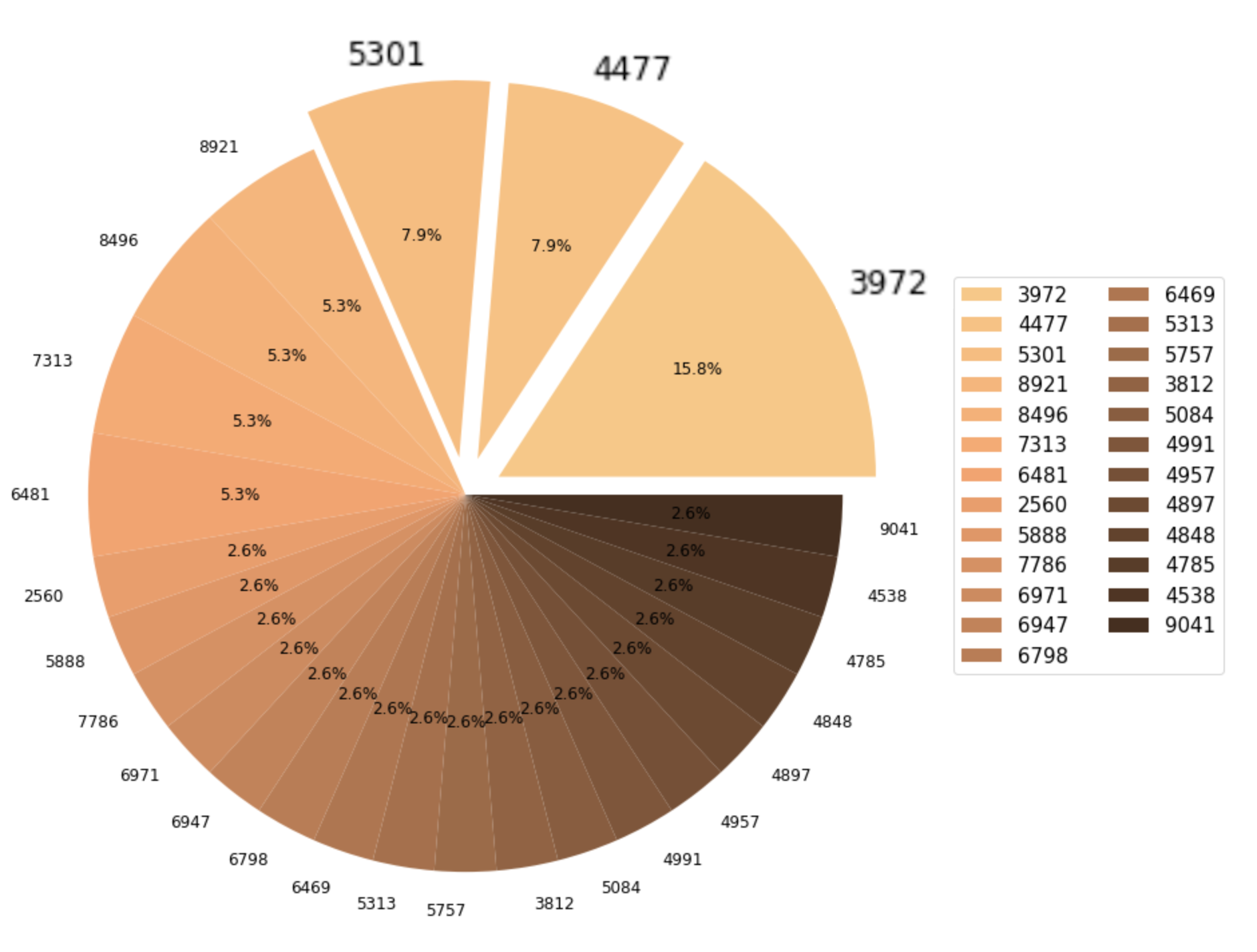}
		\label{subfig:s150}
	}
	\subfigure[Stand 138]{
		\includegraphics[width=0.3\textwidth]{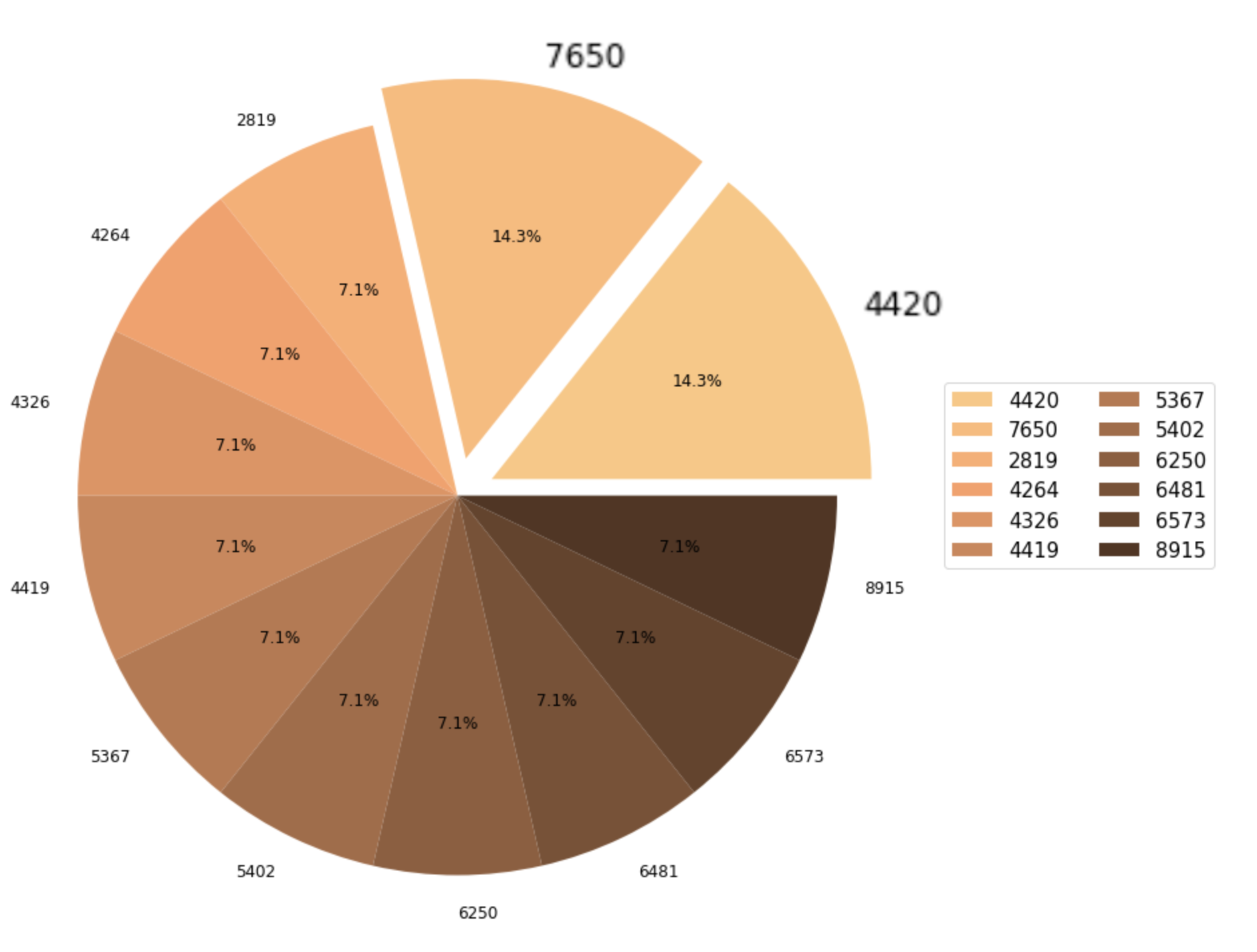}
		\label{subfig:s138}
	}
	\caption{Users returning shared E-bikes at stands with low return accuracy.}
	\label{fig:users_in_stands}
	\vspace{-0.1in}
\end{figure*}

In a similar way, we can also compare the accuracy at the level of bike stands before and after adopting the tolerance. The result is shown in \autoref{fig:fig7}, which demonstrates the number of both accurate and inaccurate return records and the return inaccuracy for several stands with and without tolerance of 50 metres, and accordingly it is obvious that the inaccuracy for some stands is remarkably higher compared with others (e.g., stand 25, 150 and 138). Thus, we conducted further analysis on users who frequently return their shared E-bikes at these stands, and the results are shown in \autoref{fig:users_in_stands}. It could be found that for different stands, the reasons for the low return accuracy might be different. As shown in Figure \ref{subfig:s25}, user 2277 contributed 74.6\% of incorrect returns at stand 25, while the other 25.4\% is attributed to other users. Thus, it is very likely that the poor return accuracy at the stand 25 is due to the inappropriate parking behaviours of some specific users entering or leaving this building frequently. In contrast, at stand 138, which locates at the Parking Lot near the intersection of ``Parnell Street'' and ``O'Connell Street Upper'', and it could be seen in Figure \ref{subfig:s138}, the number of inaccurate returns contributed by users are almost identical, and it is clear that the accurate station at ``O'Connell Street outside AIB Bank'' is very close to stand 138, so a potential reasoning could be that the low return accuracy of the stand might be related to the high traffic flow on the ``Parnell Street'', where the cyclists must move across the street if they want to return their bikes properly to the ``O'Connell Street''. Finally, for the stand 150 shown in Figure \ref{subfig:s150}, which locates closely to the ``AIB Bank on Richmond Street South'', the number of users making inappropriate returns is significantly higher compared with the other two stands. The nearest accurate return station, which locates at the location ``Richmond Row, outside Atlas Language School'', is around 200 metres away from this stand. Thus, a reasonable assumption here could be that the return accuracy at the stand 150 was affected by a lot of people who often cannot find a parking space in the designated parking area near the language school.

\begin{figure*}[ht]
	\vspace{-0.1in}
	\centering
	
	\subfigure[User 2529]{
		\includegraphics[width=0.3\textwidth]{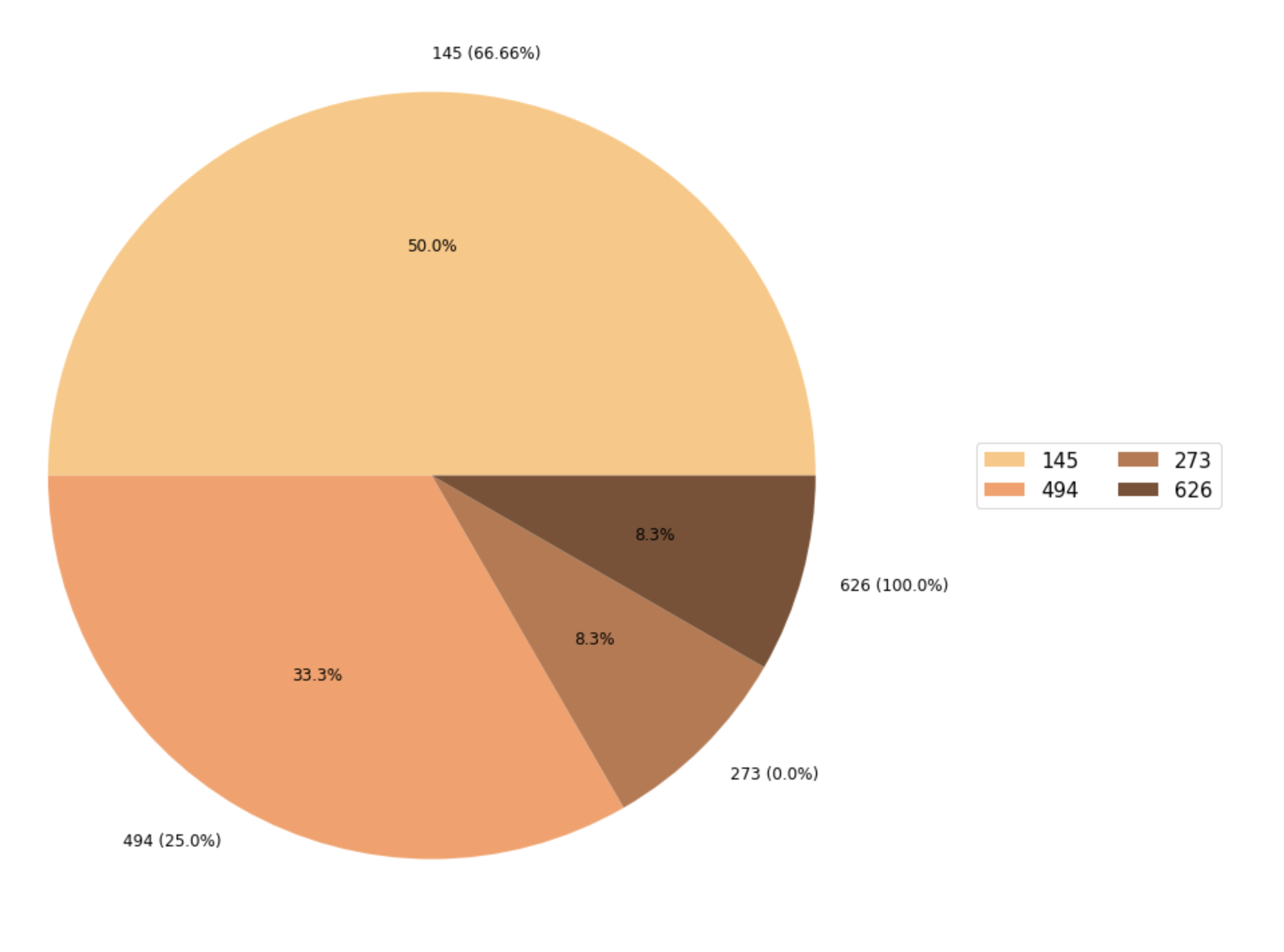}
		\label{subfig:u2529}
	}
	\subfigure[User 5035]{
		\includegraphics[width=0.3\textwidth]{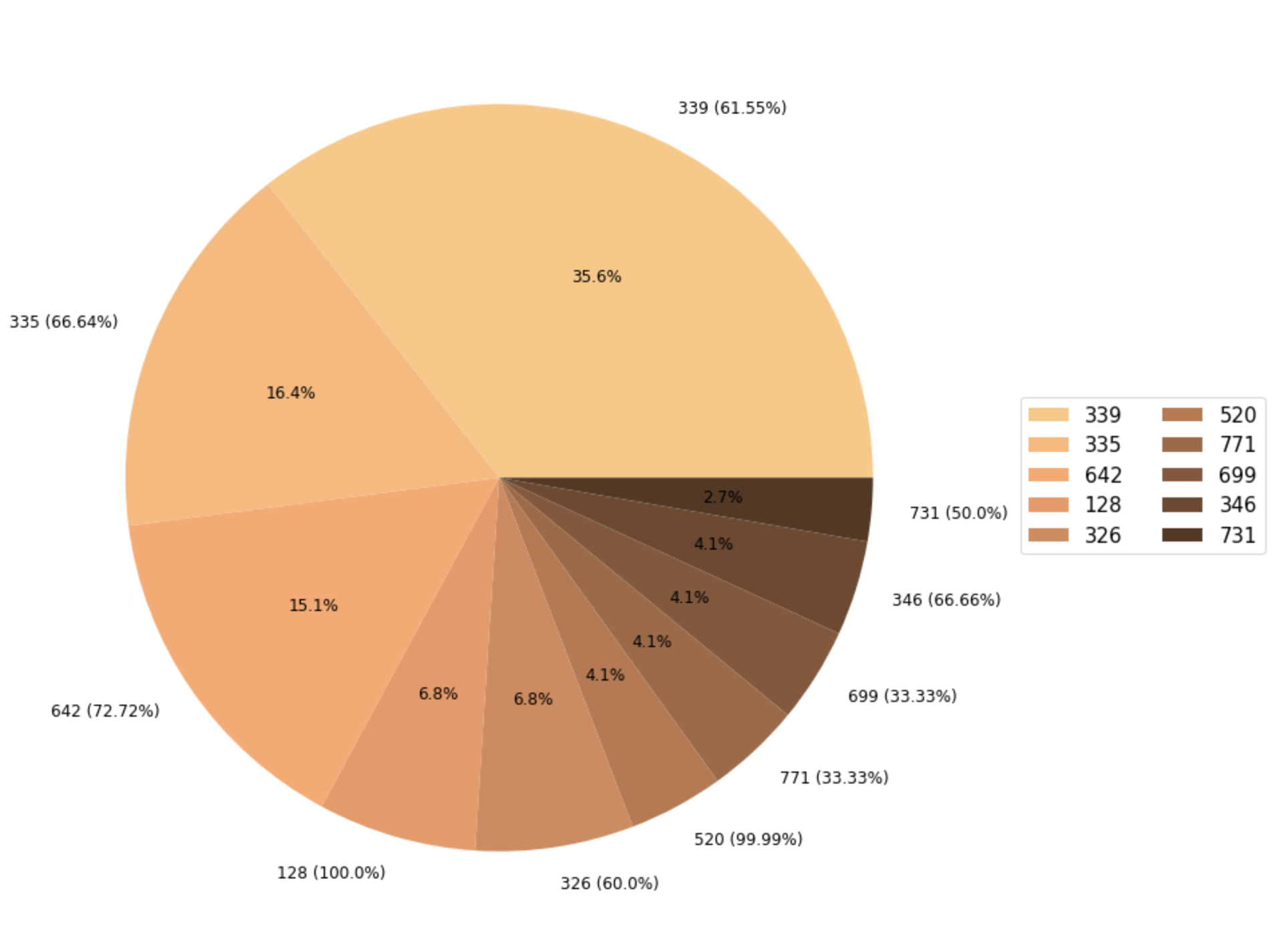}
		\label{subfig:u5035}
	}
	\subfigure[User 6263]{
		\includegraphics[width=0.3\textwidth]{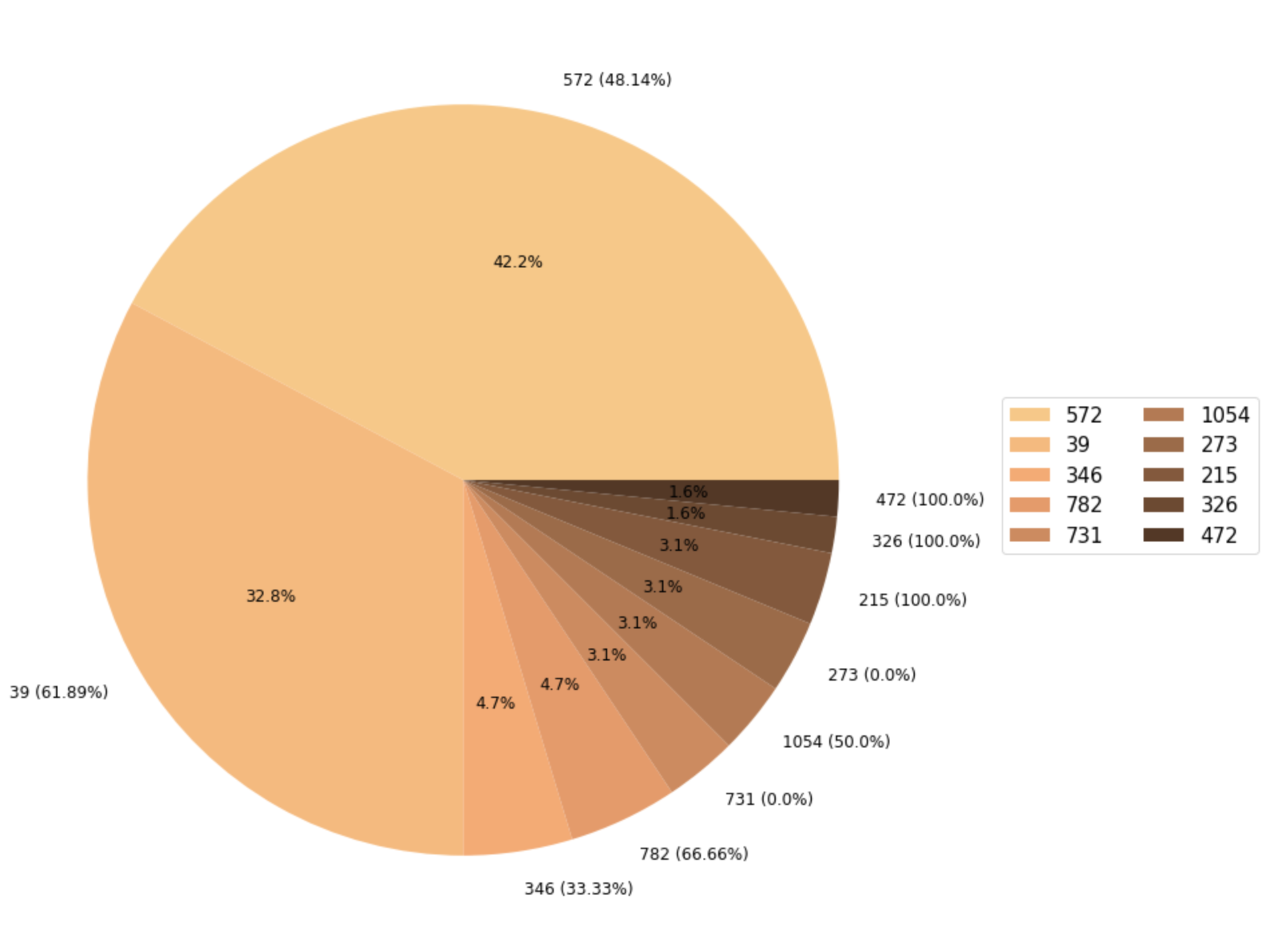}
		\label{subfig:u6263}
	}
	
	\subfigure[User 2622]{
		\includegraphics[width=0.3\textwidth]{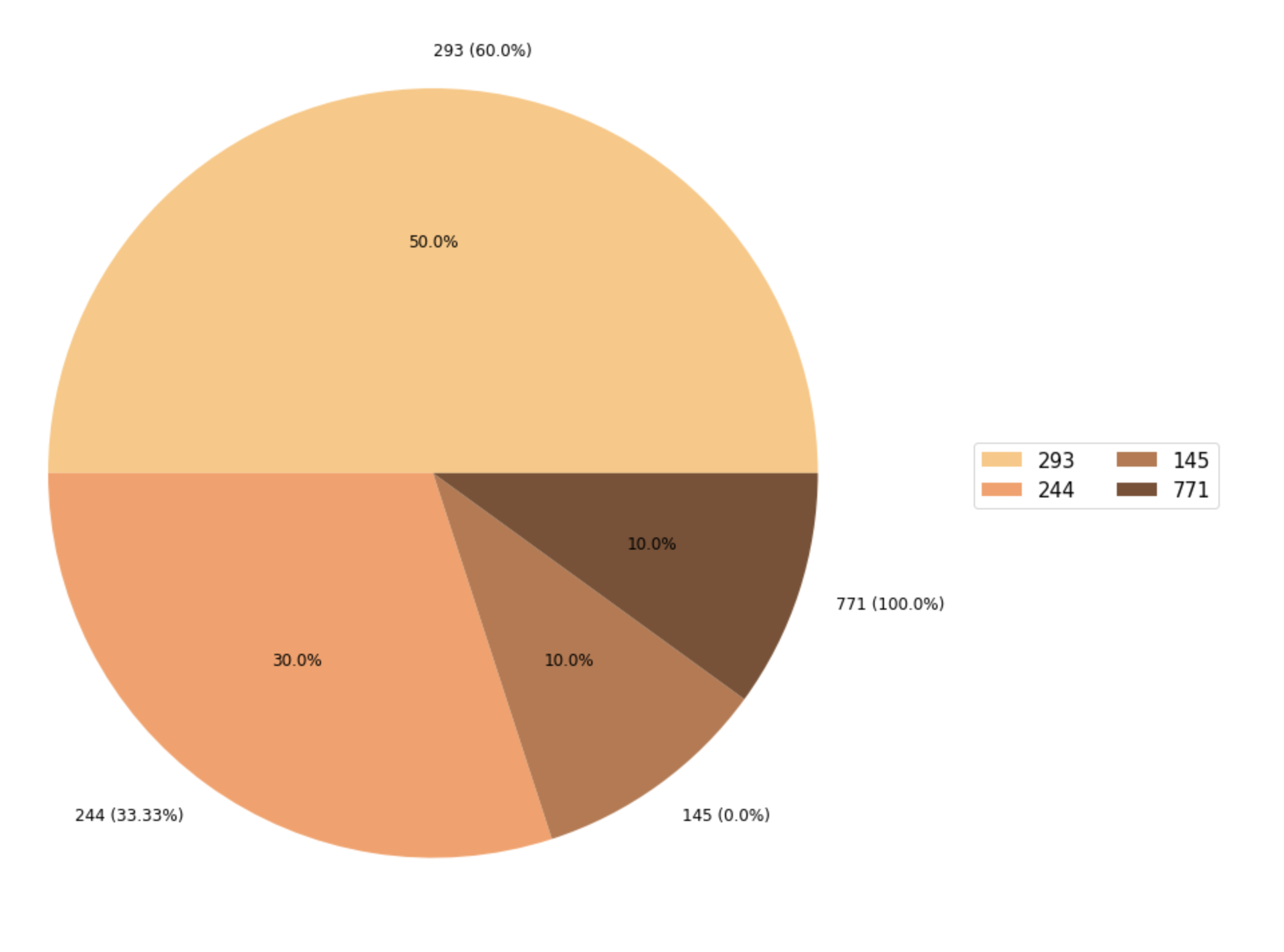}
		\label{subfig:u2622}
	}
	\subfigure[User 6020]{
		\includegraphics[width=0.3\textwidth]{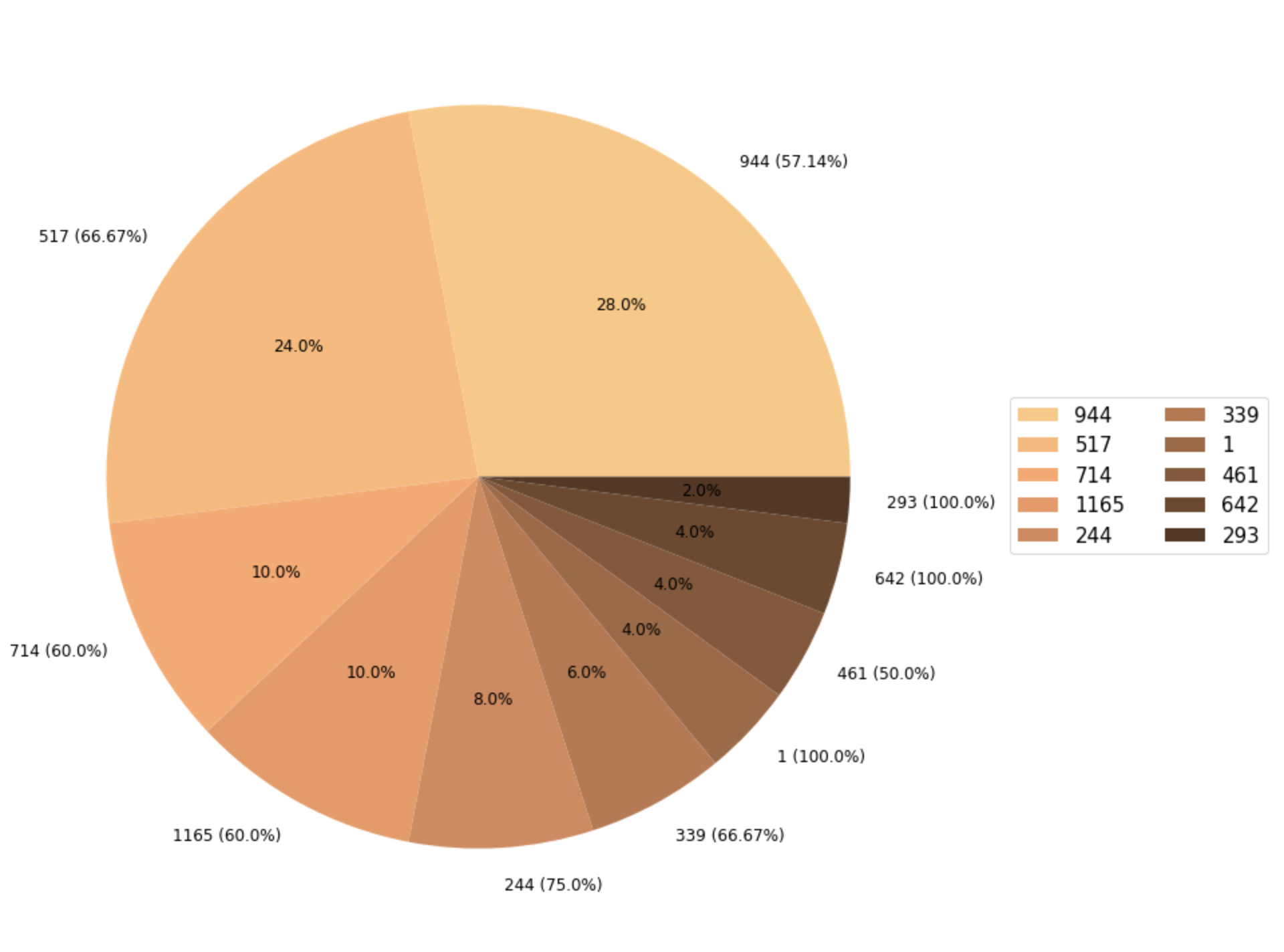}
		\label{subfig:u6020}
	}
	\subfigure[User 5382]{
		\includegraphics[width=0.3\textwidth]{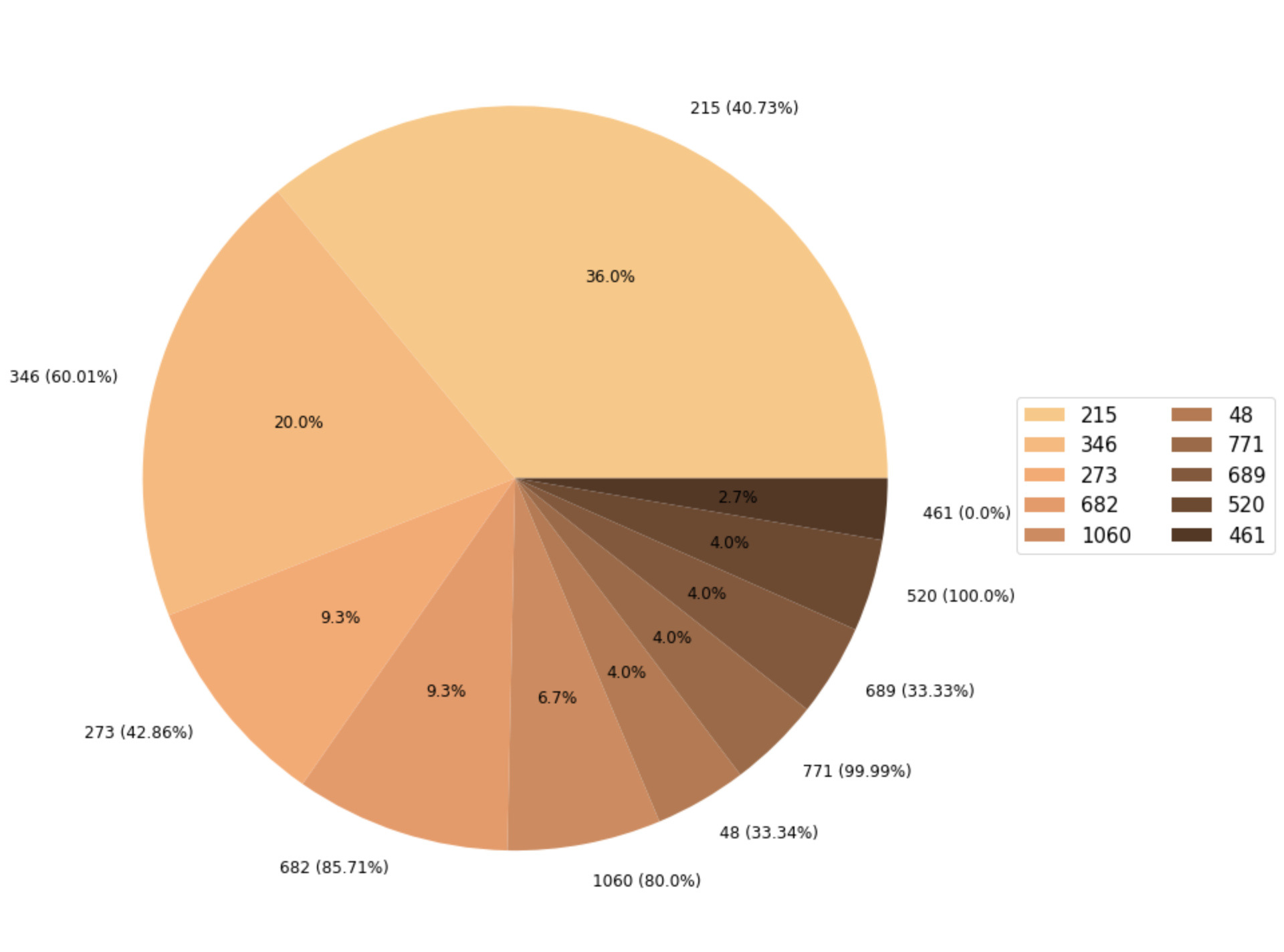}
		\label{subfig:u5382}
	}
	\caption{Performance of users with low return accuracy for each stand.}
	\label{fig:user_stand}
	\vspace{-0.1in}
\end{figure*}

\subsection{Inference from the starting point}

Up to now, we have analysed our data and reported the results primarily based on the information of return stands and their coordinates. These results are helpful for us to understand the users' parking behaviours at the end of their journeys, but these may not be good enough to infer the behaviours of users proactively. To do this, we also need to leverage the information from users' trip records based on their starting points. Mathematically, let $S:=\{1,2,\dots, n\}$ be the indexing set of stands with a capacity $n$. Let $P^{i}_{a,b}$ denote the probability for user $i$ moving from stand $a \in S$ to stand $b \in S$, then it can be calculated by:

\begin{equation} \label{ep1}
P^{i}_{a,b} = \frac{N_{a,b}^{i}}{N_{a}^{i}}\qquad
\end{equation}

\noindent where $N_{a,b}^{i}$ is the number of travel records of the user $i$ with stand $a$ as the starting point and $b$ as the ending point, and $N_{a}^{i}$ represents the total number of travel records of the user $i$ with stand $a$ as the starting point. Similarly, let $A^{i}_{a,b}$ be the return accuracy for the user $i$ with stand $a$ as the starting point and stand $b$ as the destination, then  it can be represented by the formula below:

\begin{equation} \label{ep2}
A^{i}_{a,b} = \frac{T^{i}_{a,b}}{T^{i}_{a,b} + F^{i}_{a,b}}\qquad
\end{equation}

\noindent where $T^{i}_{a,b}$ is the number of travel records with stand $a$ and $b$ as the starting and ending point when the E-bike is properly returned, while $F^{i}_{a,b}$ denotes the number of records for improperly returned trips of the user. Now, we define

\begin{equation} \label{ep3}
P^{i}_{a} := \sum_{j=1}^{n}P^{i}_{a,j} \cdot A^{i}_{a,j} \qquad \forall i 
\end{equation}
which represents the score of the user $i$, i.e., a value between 0-1, properly returns his/her E-bikes starting from the specific stand $a \in S$. Given this, we can obtain the prior probability of a user at the beginning of his/her journey. Our numerical results for the dataset are illustrated in \autoref{fig:user_stand}, where the six most inconsiderate E-bike users are summarised with respect to their respective starting stands presented in the pie charts.

\section{Conclusion and Future Work}\label{sec:conclusion}

In this paper, we have investigated the shared E-bike users' parking behaviour by using a realistic dataset provided by MOBY bike. Our results have shown that up to 12.9\% users were not parking their bikes properly with respect to the highest technical toleration of GPS devices set by MOBY. A variety of visualisation and statistical analysis on the given dataset have been included to demonstrate the users' parking behaviours as per the company's current parking management strategy from different perspective. Given the insight from the data, our future work will focus on the design of a parking assistance system to encourage those inconsiderate users to park their shared E-bikes properly. Incentive mechanisms will be designed and performance analysis for return accuracy will also be carried out as part of our future work.

\section*{Acknowledgement}

This publication has emanated from research supported in part by Science Foundation Ireland under Grant Number SFI/12/RC/2289\_P2, co-funded by the European Regional Development Fund. Our data analysis is fully anonymous and GDPR-complied. The authors would like to thank the support from the Smart DCU Projects facilitator Kieran Mahon for his assistance with data collection and delivery for our analysis. The authors would also like to thank the technical support team from MOBY for useful discussion.

\bibliographystyle{ieeetran}
\bibliography{reference}

\end{document}